\documentclass[letterpaper]{article} 
\usepackage{aaai24}  
\usepackage{times}  
\usepackage{helvet}  
\usepackage{courier}  
\usepackage[hyphens]{url}  
\usepackage{graphicx} 
\usepackage{natbib}  
\usepackage{caption} 
\usepackage{amsmath}
\usepackage{amssymb}
\usepackage{subfigure}
\usepackage{multirow}
\usepackage{float}
\usepackage[table]{xcolor}
\usepackage{makecell}
\usepackage{arydshln}
\usepackage{algorithm}
\usepackage{algorithmic}

\usepackage{newfloat}
\usepackage{listings}
\DeclareCaptionStyle{ruled}{labelfont=normalfont,labelsep=colon,strut=off} 
\floatstyle{ruled}
\newfloat{listing}{tb}{lst}{}
\floatname{listing}{Listing}
\title{Propagation Tree Is Not Deep: Adaptive Graph Contrastive Learning Approach for Rumor Detection}
\author{
    Chaoqun Cui,
    Caiyan Jia\thanks{Corresponding authors}
}
\affiliations{
    School of Computer and Information Technology \& Beijing Key Lab of Traffic Data Analysis and Mining\\
    Beijing Jiaotong University, Beijing 100044, China\\
    \{21120341,cyjia\}@bjtu.edu.cn
}

\usepackage{bibentry}

\begin{document}

\maketitle

\begin{abstract}

Rumor detection on social media has become increasingly important. Most existing graph-based models presume rumor propagation trees (RPTs) have deep structures and learn sequential stance features along branches. However, through statistical analysis on real-world datasets, we find RPTs exhibit wide structures, with most nodes being shallow 1-level replies. To focus learning on intensive substructures, we propose Rumor Adaptive Graph Contrastive Learning (RAGCL) method with adaptive view augmentation guided by node centralities. We summarize three principles for RPT augmentation: 1) exempt root nodes, 2) retain deep reply nodes, 3) preserve lower-level nodes in deep sections. We employ node dropping, attribute masking and edge dropping with probabilities from centrality-based importance scores to generate views. A graph contrastive objective then learns robust rumor representations. Extensive experiments on four benchmark datasets demonstrate RAGCL outperforms state-of-the-art methods. Our work reveals the wide-structure nature of RPTs and contributes an effective graph contrastive learning approach tailored for rumor detection through principled adaptive augmentation. The proposed principles and augmentation techniques can potentially benefit other applications involving tree-structured graphs.

\end{abstract}

\section{Introduction}

The unprecedented growth of the Internet in recent years has promoted the widespread applications of social media. Digital platforms like Weibo and Twitter have evolved into critical conduits for users to garner information and interact with each other. These platforms, while facilitating information dissemination and diverse opinion expression on a multitude of trending issues, are also breeding grounds for various rumors. Given the massive user base and the ease of use, rumors are disseminated extensively and swiftly via social media, wreaking substantial societal havoc. Therefore, there is an urgent need to establish efficacious and efficient strategies for automated rumor verification on social media.

Currently, a plethora of studies concerning rumor detection exist. Certain studies \cite{bigcn,ebgcn} have demonstrated that the propagation structure of a claim, which fully encapsulates the interrelationship between posts and harnesses the collective intelligence of the crowd, is invaluable for debunking rumors. In general, rumor detection models built upon rumor propagation structures glean discriminative features of rumors from the interrelation of comments, apprehending specific patterns of reply stances as the basis for claim classification, given that clear disparities exist between the comment stances of rumor claims and those of non-rumor claims. These models are adept at discerning these differences, which constitute one of the fundamental postulates of rumor detection methods based on rumor propagation trees \cite{rvnn}. This supposition relies, to a certain degree, on the deep structures of rumor propagation trees (RPTs). But are RPTs really deep?

Bearing this question in mind,  we undertook a statistical analysis to explore the structural characteristics of RPTs. The findings indicate that, in commonly employed rumor detection datasets and real-world social media platforms, the tree structures of claims are typically shallow. The vast majority of a claim's comments constitute 1-level replies, with the remainder primarily consisting of 2-level replies, and only a negligible number delve into deeper levels. This essentially implies that RPTs are not characterized by deep tree structures, but instead exhibit wide structures.

As per our statistical analysis, the majority of nodes in RPTs are 1-level replies, all pointing directly to the root node (i.e., the source post), which highlights the significance of the root node. Further, it is plausible that the majority of the 1-level replies, due to their lack of deeper engagement, may contain less informational value in rumor identification compared to nodes with more extensive paths. Based on these findings, we propose the Rumor Adaptive Graph Contrastive Learning (RAGCL) method. RAGCL utilizes node centrality measures to generate augmented views of RPTs and leverages graph contrastive learning methods to facilitate graph neural networks (GNNs) in learning crucial rumor discriminative features from the deep sections of RPTs. Empirical studies demonstrate the effectiveness of RAGCL.

In summary, the contributions of this study are as follows.
\begin{itemize}
\item Our statistical survey has unveiled that RPTs primarily exhibit a wide tree structure, breaking the stereotype of a deep tree structure in previous studies. This shifts the understanding of information propagation processes on social media platforms.
\item Informed by the structural characteristics of RPTs and inspired by current research on graph self-supervised learning \cite{gca}, we propose the RAGCL method to learn discriminative features for rumor detection.
\item In light of the unique tree structure of RPTs, we propose three guiding principles to be followed when designing adaptive data augmentation methods for RPTs.
\item Our experimental results underscore the superior performance of RAGCL in comparison to the current state-of-the-art (SOTA) methods and substantiate the validity of our three principles.
\end{itemize}

\section{Related Work}

In this section, we will review the related works on rumor detection and graph contrastive learning.

\subsection{Social Media Rumor Detection}

To debunk rumors, various efforts have been made. Among the existing studies, early methods mainly take advantage of traditional classification methods by using hand-crafted features \cite{dtc,rfc}. In recent years, with the advent of deep learning, more effective approaches have emerged, resulting in significant improvements in rumor detection performance. These approaches can be broadly categorized into four classes, including time-series based methods \cite{weibo,yucnn,liuandwu} which model text content or user profiles as time series, propagation structure learning methods \cite{rvnn,bigcn,ebgcn,gacl,adgscl,urumor,qyh,pep} which consider the propagation structures of source rumors and their replies, multi-source integation methods \cite{ms1,ms2} which combine multiple resources of rumors including post content, user profiles, heterogeneous relations between posts and users, multi-modal fusion methods \cite{eann,otherrumor1,vga} which incorporate both post content and related images to effectively debunk rumors. 

Although large language models have achieved excellent performance in many natural language processing tasks \cite{gpt3,rlhf,sspo}, propagation structure information remains critically important in the field of rumor detection. Numerous SOTA models bank on learning the representations of RPTs utilizing GNNs. \citet{rvnn} designed a bottom-up and top-down tree-structured recursive neural network to extract information from RPTs. In a similar vein, \citet{bigcn} applied a bidirectional GCN alongside a root node feature enhancement technique to address rumor detection tasks. Furthermore, \citet{gacl} incorporated contrastive loss with adversarial training to learn representations robust to rumor noise. These studies bear testament to the efficacy of propagation structure learning in accurately identifying rumors.

\subsection{Graph Contrastive Learning}

The advancement of deep learning has instigated progress across numerous studies predicated on neural message passing algorithms \cite{messagepass}. These algorithms learn graph representations in a supervised manner and have attained SOTA results across a wide array of tasks \cite{messagepass1,messagepass2,messagepass3}. 
In recent years, graph self-supervised learning methods have gradually emerged to leverage unlabeled data for addressing the problem of scarce labeled data, with most of them being graph contrastive learning methods.
Contrastive learning methods have been widely applied in the domain of image representation learning \cite{moco,simclr}, and subsequently extended to the realms of text \cite{declutr,deepchannel,cert} and graph data \cite{dgi,infograph}. Graph contrastive learning methodologies have evolved from initial methods premised on mutual information maximization \cite{dgi,infograph} to contemporary methods based on graph augmentation \cite{mvgrl,graphcl,joao}. 
Graph contrastive learning methods based on graph augmentation first employ diverse graph augmentation strategies (such as node drop, edge perturbation, etc.) to acquire varying views of a given graph, thereafter constructing positive and negative samples in the contrastive loss. Ultimately, graph representations are learned by minimizing the contrastive loss. 
To accommodate different types of graph datasets, several research studies have focused on adaptive graph augmentation \cite{gca,joao,autogcl}. RAGCL represents a novel adaptive graph augmentation approach for rumor detection to learn robust and discriminative representations of RPTs.

\section{Analysis on Propagation Tree}

\begin{figure}[t]
  \centering
  \subfigure[False-Rumor]{\includegraphics[width=0.48\linewidth]{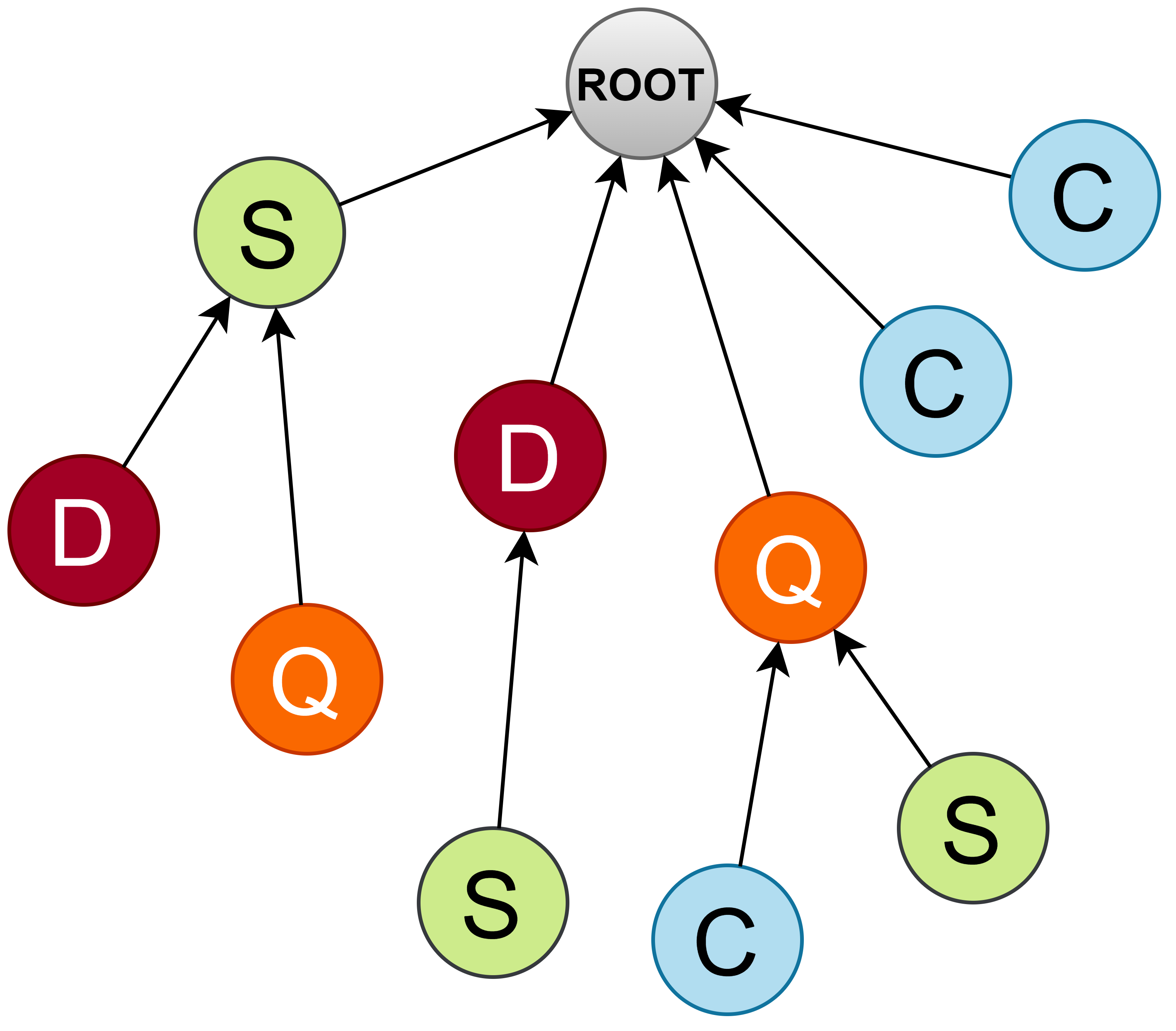}}
  \subfigure[True-Rumor]{\includegraphics[width=0.48\linewidth]{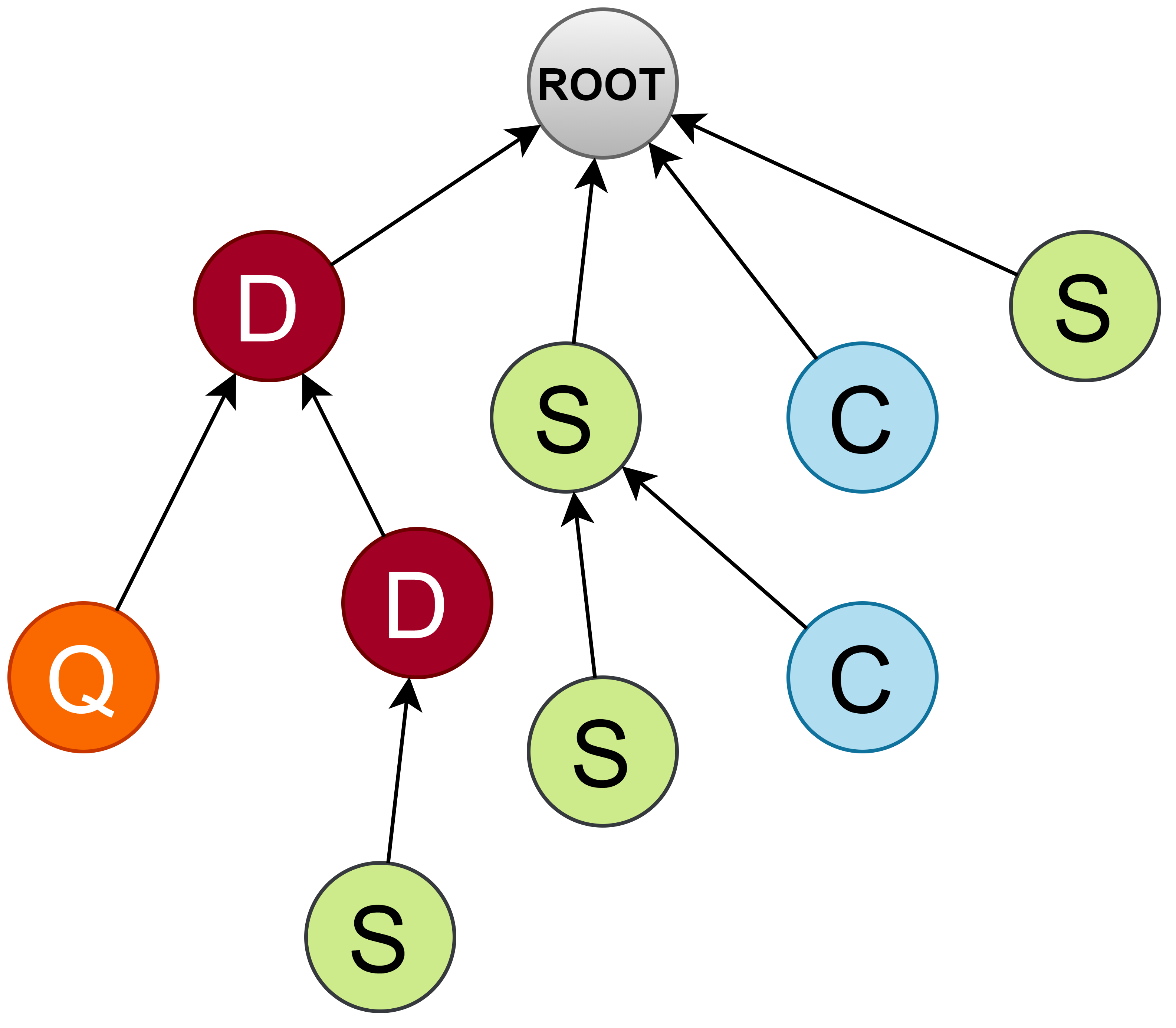}}
  \caption{The stances in rumor propagation trees.}
  \label{fig:st}
\end{figure}

As shown in Figure~\ref{fig:st}, current propagation structure learning based rumor detection methods are devoted to collect support (S), deny (D), question (Q), comment (C) and other stances between a reply and its source post, and pairs of replies \cite{rvnn}. For different classes of claims, there exist noticeable differences in their stance patterns, which can serve as discriminative features for rumor identification. For instance, the true stance of a D-D relation is S, whereas the true stance of a D-S relation is D.
Current propagation structure learning methods exploit stance features between sequential nodes on the same branch of the tree structure for rumor detection \cite{bigcn,gacl}. Nonetheless, these features are dependent on the depth structure of RPTs. But, are RPTs truly deep?

In the present study, we conduct a thorough exploration of the structural properties inherent to RPTs, deploying a statistical approach. The datasets under scrutiny consist of Weibo \cite{weibo}, DRWeibo\footnote{\url{https://github.com/CcQunResearch/DRWeibo}}, Twitter15, and Twitter16 \cite{twitter1516}. Further, we survey two large-scale unlabeled public datasets, namely UWeibo\footnote{\url{https://github.com/CcQunResearch/UWeibo}} and UTwitter\footnote{\url{https://github.com/CcQunResearch/UTwitter}}. Data from these datasets originate from popular posts on Weibo and Twitter platforms, mirroring the universal traits of claims within social media environments. The statistical results are shown in Table~\ref{tab:structure}. The entries beneath the dotted line in Table~\ref{tab:structure} denote the mean count of replies, 1-level replies, 2-level replies, deeper ($>$2) replies, and 1-level replies with subsequent replies per claim in the dataset, respectively. The statistics lead us to the ensuing conclusions.
\begin{itemize}
\item \textbf{RPTs resemble wide trees rather than deep ones.} 1-level replies constitute the majority of all replies within RPTs, with proportions of 65.1\%, 77.8\%, 70.7\%, and 64.2\% for the four labeled datasets respectively.
\item \textbf{Only a minimal portion of 1-level replies within RPTs spawn subsequent replies.} Amongst all 1-level replies in RPTs, only a fraction give rise to further replies, with percentages of 9.7\%, 6.4\%, 10.4\%, and 10.8\%.
\item \textbf{Deep replies within a RPT are seldom observed.} Deep replies make up a tiny fraction of all replies in RPTs, with percentages of 13.8\%, 4.4\%, 17.3\%, and 23.4\%. This infers that the model is constrained to learning the aforesaid stance features from a limited set of replies.
\end{itemize}

\begin{table*}[h]
\centering
\begin{tabular}{ccccccc}
\Xhline{1.0pt}
\rowcolor{gray!20}
\textbf{Statistic} & \textbf{Weibo} & \textbf{DRWeibo} & \textbf{Twitter15} & \textbf{Twitter16} & \textbf{UWeibo} & \textbf{UTwitter}\\
\hline
\textbf{language} & zh & zh & en & en & zh & en \\
\textbf{\# claims} & 4664 & 6037 & 1490 & 818 & 209549 & 204922\\
\textbf{\# non-rumors} & 2351 & 3185 & 374 & 205 & - & -\\
\textbf{\# false rumors} & 2313 & 2852 & 370 & 205 & - & -\\
\textbf{\# true rumors} & - & - & 372 & 207 & - & -\\
\textbf{\# unverified rumors} & - & - & 374 & 201 & - & -\\
\hdashline
\textbf{\# avg reply} & 803.5 & 61.8 & 50.2 & 49.1 & 50.5 & 82.5\\
\textbf{\# avg 1-level reply} & 522.9(65\%) & 48.1(78\%) & 35.5(71\%) & 31.6(64\%) & 36.4(72\%) & 48.5(59\%)\\
\textbf{\# avg 2-level reply} & 169.3(21\%) & 11.0(17\%) & 5.9(12\%) & 6.0(12\%) & 10.2(20\%) & 21.5(26\%)\\
\textbf{\# avg deeper reply} & 111.2(14\%) & 2.7(5\%) & 8.7(17\%) & 11.5(24\%) & 4.0(8\%) & 12.5(15\%)\\
\textbf{\# avg responded 1-level reply} & 50.7 & 3.1 & 3.7 & 3.4 & 4.1 & 8.0\\
\Xhline{1.0pt}
\end{tabular}
\caption{Statistics of the datasets.}
\label{tab:structure}
\end{table*}

\begin{figure}[t]
  \centering
  \includegraphics[width=8cm]{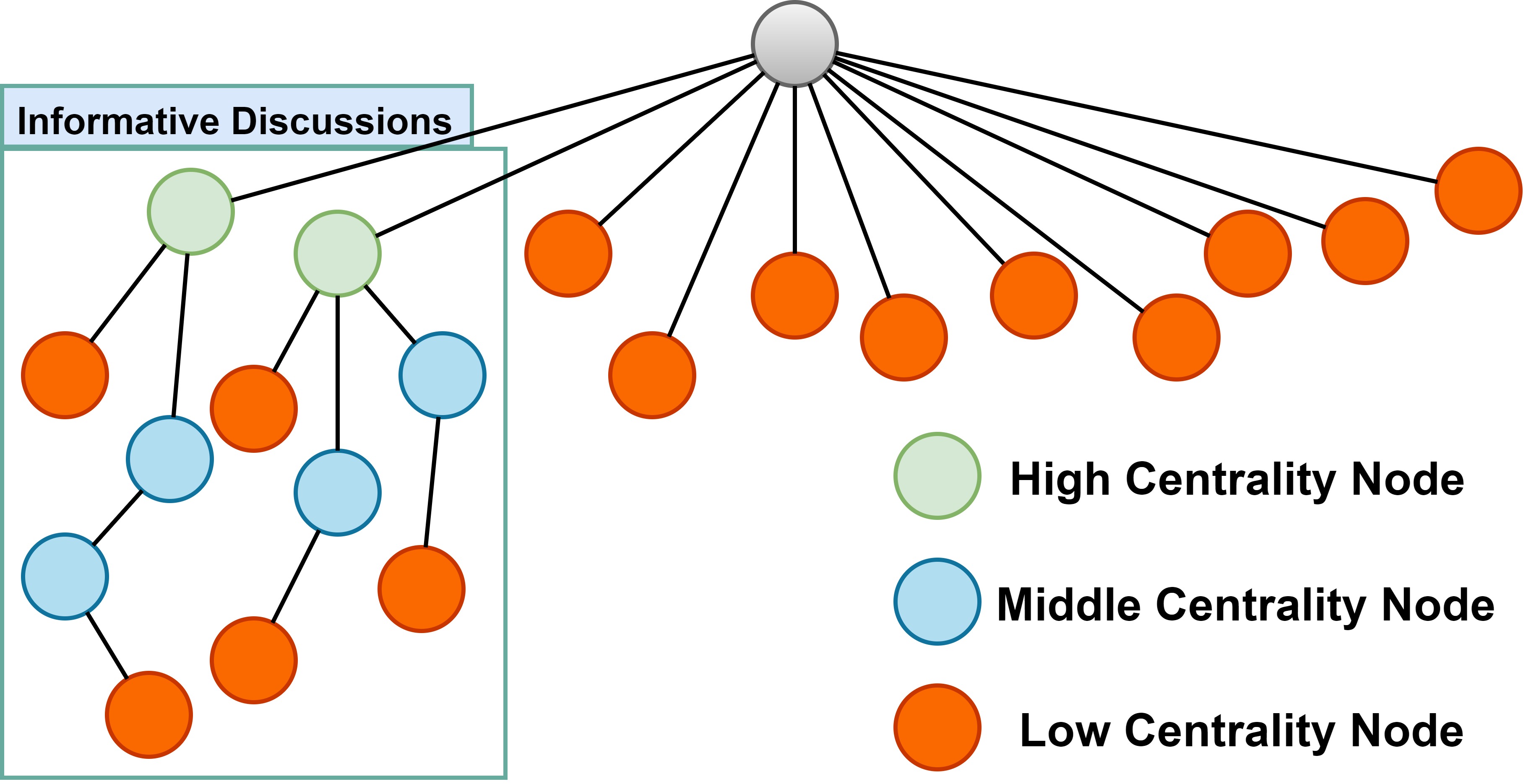}
  \caption{A rumor propagation tree.}
  \label{fig:rpt}
\end{figure}

Both UWeibo and UTwitter datasets also exhibit these three characteristics, signifying that these traits are pervasive attributes of claims on social media platforms. The above observations paint the generalized structure of a RPT, as illustrated in Figure~\ref{fig:rpt}. It's noticeable that only nodes enclosed within the box in Figure~\ref{fig:rpt} carry the aforementioned stance features, while most of the nodes in the tree are 1-level nodes without further reply (no deep structure). 

Based on the above observations, we can conceptualize a RPT as a highly imbalanced graph, with imbalances reflected in the following two aspects:
\begin{itemize}
\item The root node of a RPT features highly dense connections, whereas connections at the remaining nodes are exceedingly sparse.
\item The intensive discussions and informative portions of a RPT are predominantly found within a limited number of 1-level replies (the two green nodes in Figure~\ref{fig:rpt}). In contrast, the majority of the 1-level replies that lack further deeper responses also lack discriminative features that can aid in rumor identification.
\end{itemize}

Such characteristics are determined by users' habits of using social media and the order in which platforms display comments. In general, users are inclined to reply directly to source posts rather than to other users' comments. Additionally, platforms such as Weibo and Twitter tend to sort replies based on popularity rather than the chronological order of posting. This contributes to the imbalance in the information distribution within a propagation tree. 
With the aim of enhancing our model's focus on the intense and informative discussions of RPTs and reducing the influence of a large number of unresponded 1-level replies, we put forward our RAGCL method. The objective of RAGCL is to stress the importance of comments within RPTs that have intensive replies, while also focusing on root nodes by directing the aggregation of information from other nodes towards these roots, considering the wide structures of RPTs.

\section{Method}

We will present the design of RAGCL in this section.

\subsection{Notation}

The rumor detection task can be defined as a graph-level classification task. Specifically, we denote a labeled claim dataset as $\mathbb{C}=\left \{c_{1},c_{2},\cdots ,c_{m}\right \}$, where $c_{i}$ represents the $i$-th claim and $m$ represents the number of labeled claims. Each labeled claim $c=(y,G)$ consists of its ground-truth label $y\in \left \{N,R\right \}$ (i.e., Non-rumor or Rumor) or fine-grained label $y\in \left \{N,F,T,U\right \}$ (i.e., Non-rumor, False Rumor, True Rumor, Unverified Rumor) and its propagation structure $G=(V,E)$, where $V$ and $E$ represent the set of nodes (a source post and comments of the claim) and edges (the relations between pairs of replies or source post and a reply), respectively. The set of propagation structure graphs corresponding to all claims is $\mathbb{G}=\left \{G_{1},G_{2},\cdots ,G_{m}\right \}$. 
The goal of rumor detection task is to learn a classifier $f:\mathbb{G}\longrightarrow Y$ 
$(Y=\{y_1,y_2\cdots y_m\})$ from dataset $\mathbb{C}$.

\subsection{Framework}

\begin{figure*}[t]
  \centering
  \includegraphics[width=\textwidth]
  {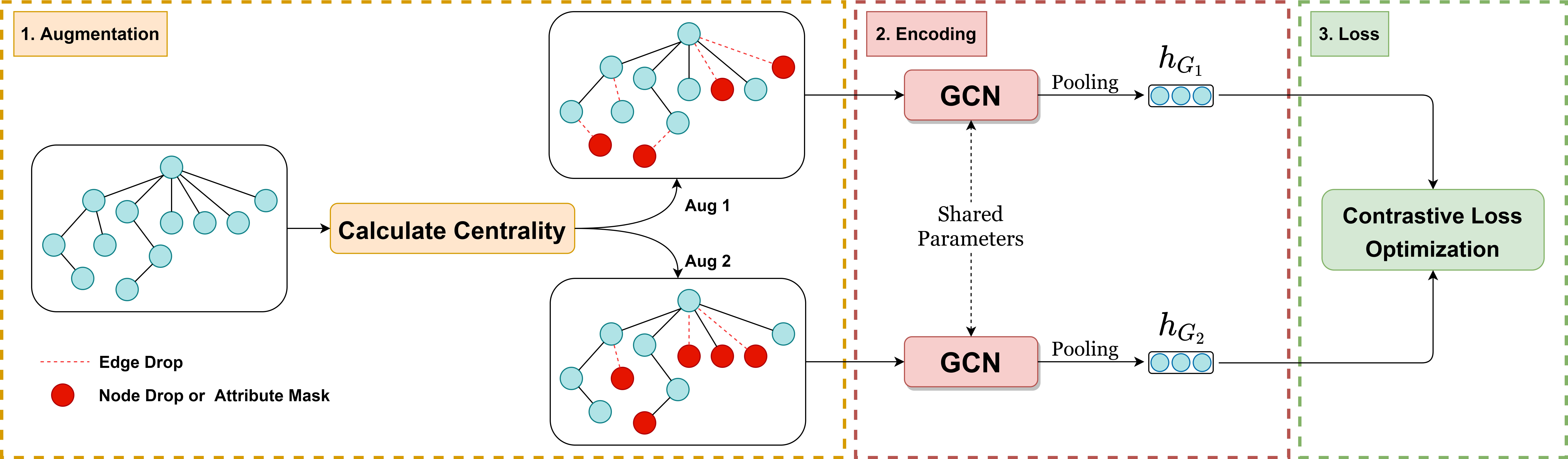}
  \caption{The framework of RAGCL.}
  \label{fig:framework}
\end{figure*}

From the aforementioned analysis, it is important to learn discriminative features from nodes with deep structures (e.g., the nodes in the box in Figure~\ref{fig:rpt}). These nodes and their corresponding edges possess a conspicuously higher importance compared to the nodes located outside the box. Based on this idea, we introduce RAGCL, an adaptive graph contrastive learning framework purposefully engineered for rumor detection. RAGCL assigns varying levels of importance to nodes and edges within a RPT based on a selected node centrality measure. Subsequently, varying probabilities of drop or mask, informed by these scores, are employed to adaptively generate two graph augmented views of the RPT, utilizing node drop, attribute mask, or edge drop operators. The contrastive loss is subsequently minimized to learn the tree's representation. A comprehensive illustration of the RAGCL process is presented in Figure~\ref{fig:framework}.

\subsection{Augmentation Principle}

Node centrality is an index to measure the importance of nodes in a graph. There are three recommended node centrality measures used by RAGCL, including degree centrality \cite{degree}, betweenness centrality \cite{betweenness} and PageRank centrality \cite{pagerank}. 
\begin{itemize}
\item \textbf{Degree centrality} takes the degree of nodes as the measure of node centrality. The idea is that a post with multiple replies is important in a RPT. RAGCL uses the node out-degree of top-down graph of a RPT as the measure of degree centrality.
\item \textbf{Betweenness centrality} calculates all shortest paths of any two nodes in a graph. A node becomes prominent in terms of betweenness centrality if a multitude of these paths transit through it. RAGCL utilizes either top-down or bottom-up graphs to ascertain betweenness centrality.
\item \textbf{PageRank centrality} is commonly used in web page ranking. Its basic idea is that the importance of a page on Internet depends on the quantity and quality of inbound links. RAGCL leverages the bottom-up graph of a RPT to compute PageRank centrality.
\end{itemize}

Given our prior analysis on the structural characteristics of RPTs, we have summarized the following three principles for assigning importance scores to nodes and edges.
\begin{itemize}
\item \textbf{Principle 1}: Given the pivotal role of source posts \cite{bigcn,gacl}, the root nodes of RPTs are exempt from the data augmentation procedure.
\item \textbf{Principle 2}: Nodes and edges with deep replies within RPTs (referenced within the boxed portion of Figure~\ref{fig:rpt}) should be preserved to the greatest extent feasible. 
\item \textbf{Principle 3}: In the deep parts of RPTs, the nodes in low-level should be retained in data augmentation more than its deeper successor nodes, because the successor nodes are basically discussed around their parent nodes, so they should hold relatively lower importance. 
\end{itemize}

Other node centrality measures, such as eigenvector centrality \cite{eigenvector}, Katz centrality \cite{katz}, and closeness centrality \cite{closeness}, are deemed unsuitable for RPTs due to inherent characteristics which preclude adherence to the aforementioned principles. The node colors in Figure~\ref{fig:rpt} show the magnitude of node centrality that should be obtained according to the above principles. Furthermore, to ensure compliance with Principle 2, RAGCL assigns the root node of a RPT the minimum value from among all node centralities within the graph, given its dense characteristic. Throughout the data augmentation process, the importance of an edge in RAGCL is gauged by the centrality of the edge's two constituting nodes. An excessively high root node centrality could artificially inflate the importance of edges connecting the root node with unresponded 1-level replies, thereby contravening Principle 2.

\subsection{Adaptive Graph Augmentation}

RAGCL conducts adaptive data augmentation according to node centrality, yielding two augmented views of the RPT. It primarily utilizes three unique data augmentation operators: node dropping, attribute masking, and edge dropping. During the training phase, two out of these three operators are selected. RAGCL employs node centrality to assign importance scores to nodes and edges, after which it computes the probability of dropping or masking for data augmentation.

\subsubsection{Node Dropping}

Consider a propagation graph $G$ of any given claim within dataset $\mathbb{C}$. Given a node centrality measure $\varphi _{c}(\cdot ):\mathcal{V}\rightarrow \mathbb{R}^{+}$, $\mathcal{V}$ is the space where node $v$ is located, and the final node centrality value of node $v$ is represented by $\varphi _{c}(v)$. Node dropping in RAGCL involves assigning a drop probability $p_{v}^{n}$ to each node $v$ and removing a portion of nodes (along with the edges connected to these nodes) from the node set $V$ in accordance with this probability to yield an augmented view. It is noteworthy that the root node is never dropped in this operation. The node importance score $w_{v}^{n}$ is set as node centrality value, that is, $w_{v}^{n}=\varphi _{c}(v)$. Given that the value of node centrality might vary across several orders of magnitude, $s_{v}^{n}=log\, w_{v}^{n}$ is set to alleviate the influence of densely connected nodes. The node drop probability is derived following the subsequent normalization procedure.
\begin{equation}
p_{v}^{n}=\frac{s_{max}^{n}-s_{v}^{n}}{s_{max}^{n}-u_{s}^{n}}\cdot p_{n},
\end{equation}
where $p_{n}$ is a hyperparameter governing the overall probability of node dropping, and $s_{max}^{n}$ and $u_{s}^{n}$ represent the maximum and mean values of $s_{v}^{n}$, respectively.

\subsubsection{Attribute Masking}

Attribute masking in RAGCL is defined as substituting the feature vectors of a fraction of the nodes in $V$ with a zero vector. The root node is exempt from this operation. Attribute masking does not entail node removal; hence, edges connected to masked nodes are retained. The mask probability for a node $v$ is also $p_{v}^{n}$.

\subsubsection{Edge Dropping}

We adopt the top-down graph of RPTs in RAGCL. Edge dropping involves setting a drop probability $p_{uv}^{e}$ for each edge $(u,v)$, subsequently utilizing this probability to remove certain edges from the edge set $E$ to produce an augmented view. $p_{uv}^{e}$ should reflect the edge's importance, implying that the $p_{uv}^{e}$ of an essential edge should be lower than that of a less critical edge. Note that the centrality of a root node is assigned the minimum centrality value among all nodes in a graph. The importance score $w_{uv}^{e}$ is defined as the mean centralities of its two connecting nodes.
\begin{equation}
w_{uv}^{e}=(\varphi _{c}(u)+\varphi _{c}(v))/2.
\end{equation}

The drop probability is then derived based on the importance score of edge $(u,v)$. Analogously, we set $s_{uv}^{e}=log\, w_{uv}^{e}$ to mitigate the impact of densely connected nodes. The probability is then ascertained similarly as follows.
\begin{equation}
p_{uv}^{e}=\frac{s_{max}^{e}-s_{uv}^{e}}{s_{max}^{e}-u_{s}^{e}}\cdot p_{e},
\end{equation}
where $p_{e}$ is a hyperparameter utilized to regulate the overall probability of edge dropping, and $s_{max}^{e}$ and $u_{s}^{e}$ represent the maximum and mean values of $s_{uv}^{e}$, respectively.

\subsection{Contrastive Loss Optimization}

The data augmentation of a propagation graph $G$ yields two augmented views, namely $G_{1}$ and $G_{2}$. These views are processed through a GCN \cite{gcn} encoder to obtain two representations: $h_{G_{1}}$ and $h_{G_{2}}$. Within RAGCL, the unsupervised contrastive loss on the graph set $\mathbb{G}$, corresponding to the dataset $\mathbb{C}$, is formulated as follows:
\begin{equation}
\begin{split}
\mathcal{L}_{unsup}=&-\mathbb{E}_{\mathbb{P}}[sim(h_{G_{1}},h_{G_{2}})]\\
&+\mathbb{E}_{\mathbb{P}}[log(\mathbb{E}_{\tilde{\mathbb{P}}}exp(sim(h_{G_{1}},h_{G_{2}}^{'})))\\
&+log(\mathbb{E}_{\tilde{\mathbb{P}}}exp(sim(h_{G_{1}}^{'},h_{G_{2}})))],
\end{split}
\end{equation}
where $\mathbb{P}$ denotes the distribution adhered to by $\mathbb{G}$; $G$ represents an input sample drawn from $\mathbb{P}$; $G^{'}$ is a negative sample drawn from $\tilde{\mathbb{P}}=\mathbb{P}$; $sim(x_{1},x_{2})=x_{1}^{T}x_{2}/||x_{1}||\: ||x_{2}||$  is the cosine similarity.

RAGCL employs $\mathcal{L}_{unsup}$ as the regularization term of the supervised loss $\mathcal{L}_{sup}$ (calculated by $h_{G}$), and optimizes the following loss function during the training phase.
\begin{equation}
\mathcal{L}=\mathcal{L}_{sup}+\lambda \cdot \mathcal{L}_{unsup},
\end{equation}
where $\lambda$ is an tunable hyperparameter.

\section{Experiments}

In this section, we present main experimental results. Experiments on the effects of hyperparameters are in the supplementary material. 

\subsection{Experimental Configuration}

We conducted experiments on four real-world benchmark datasets, Weibo, DRWeibo, Twitter15 and Twitter16, to evaluate RAGCL's performance. Weibo and DRWeibo are Chinese binary classification datasets, and Twitter15 and Twitter16 are English multiple classification datasets. Table~\ref{tab:structure} shows the statistics of the datasets. 

We make comparisons with the following baselines.

\textbf{PLAN} \cite{plan} is based on Transformer. Its StA-PLAN version incorporates RPT structural information.

\textbf{BiGCN} \cite{bigcn} leverages two GCN encoders, a top-down and a bottom-up, and root node feature enhancement strategy to classify rumor.

\textbf{UDGCN} is a variant of BiGCN, it takes undirected graph of a RPT as the model input, and only one GCN encoder that applies root node feature enhancement strategy is used.

\textbf{GACL} \cite{gacl} performs rumor classification based on contrastive learning and adversarial training.

\textbf{DDGCN} \cite{ddgcn} can model multiple types of information in one unified framework.

The experiment setting details will be explained in the supplementary material. The experimental results are the average results of 10 random split of the datasets. 
We report the best performance of RAGCL that can be achieved with different node centrality and data augmentation combination. The source code of RAGCL is available at \url{https://github.com/CcQunResearch/RAGCL}.

\subsection{Results and Discussion}

Results in Table~\ref{tab:weibo} and~\ref{tab:twitter} show that RAGCL outperforms the baselines on all datasets. 
PLAN performs relatively poorly on all datasets and consumes more GPU resources due to Transformer architecture, which points to the necessity of adopting GNN architecture. 
BiGCN is a typical model built on the deep structure of RPT, which presupposes that the information flow in RPTs presents as a top-down propagation and a bottom-up dispersion process. However, our research findings indicate that the RPT actually manifests as a wide structure. This suggests that, for tree structures like RPT, in addition to the depth-directional information flow, the imbalanced distribution of information in the width direction is also an important characteristic, which is currently overlooked by existing techniques. 
Although GACL uses BERT \cite{bert} to extract initial feature vectors, it does not improve significantly over other baselines. This may suggest that rumor detection models are insensitive to the way initial features are extracted, and what is more crucial is the high-level model's ability to learn the interactions between nodes. Additionally, GACL utilizes supervised contrastive learning to learn the claim representation, while RAGCL, which adopts unsupervised contrastive loss, also achieves superior performance. The application of unsupervised loss allows the model to learn good representations without relying on labels. This suggests that it is feasible to use RAGCL to further enhance the rumor detection capability of the model by pretraining on large-scale, unlabeled dataset from social media platforms (such as UWeibo and UTwitter). We leave this for future research.

\begin{table*}[!h]
\centering
\begin{tabular}{cccccccccc}
 \Xhline{1.0pt}
 \rowcolor{gray!20}
 ~ & ~ & \multicolumn{4}{c}{\textbf{Weibo}} & \multicolumn{4}{c}{\textbf{DRWeibo}}\\
 \cline{3-10}
 \rowcolor{gray!20}
 \multirow{-2}{*}{\textbf{Method}} & \multirow{-2}{*}{\textbf{Class}} & \textbf{Acc.} & \textbf{Prec.} & \textbf{Rec.} & \textbf{F1} & \textbf{Acc.} & \textbf{Prec.} & \textbf{Rec.} & \textbf{F1}\\
 \hline
 \multirow{2}{*}{PLAN} & R & \multirow{2}{*}{0.915\small ±0.007} & 0.908 & 0.923 & 0.915 & \multirow{2}{*}{0.788\small ±0.005} & 0.786 & 0.760 & 0.771 \\
 & N & ~ & 0.923 & 0.907 & 0.914 & ~ & 0.793 & 0.813 & 0.802 \\
 \hline
 \multirow{2}{*}{BiGCN} & R & \multirow{2}{*}{0.942\small ±0.008} & 0.919 & 0.968 & 0.942 & \multirow{2}{*}{0.866\small ±0.010} & 0.869 & 0.849 & 0.858 \\
 & N & ~ & 0.967 & 0.918 & 0.942 & ~ & 0.863 & 0.882 & 0.872 \\
 \hline
 \multirow{2}{*}{UDGCN} & R & \multirow{2}{*}{0.940\small ±0.007} & 0.914 & 0.971 & 0.942 & \multirow{2}{*}{0.861\small ±0.010} & 0.839 & 0.871 & 0.855 \\
 & N & ~ & 0.969 & 0.910 & 0.938 & ~ & 0.882 & 0.852 & 0.867 \\
 \hline
 \multirow{2}{*}{GACL} & R & \multirow{2}{*}{0.938\small ±0.006} & 0.936 & 0.940 & 0.938 & \multirow{2}{*}{0.870\small ±0.009} & 0.865 & 0.856 & 0.860 \\
 & N & ~ & 0.940 & 0.936 & 0.938 & ~ & 0.874 & 0.882 & 0.878 \\
 \hline
 \multirow{2}{*}{DDGCN} & R & \multirow{2}{*}{0.948\small ±0.004} & 0.924 & \textbf{0.979} & 0.951 & \multirow{2}{*}{0.878\small ±0.005} & 0.872 & 0.864 & 0.868 \\
 & N & ~ & \textbf{0.976} & 0.917 & 0.946 & ~ & 0.883 & 0.891 & 0.887 \\
 \hline
 \multirow{2}{*}{\textbf{RAGCL}} & R & \multirow{2}{*}{\textbf{0.962\small ±0.005}} & \textbf{0.956} & 0.968 & \textbf{0.962} & \multirow{2}{*}{\textbf{0.894\small ±0.004}} & \textbf{0.893} & \textbf{0.877} & \textbf{0.885} \\
 & N & ~ & 0.969 & \textbf{0.957} & \textbf{0.963} & ~ & \textbf{0.895} & \textbf{0.909} & \textbf{0.902} \\
 \Xhline{1.0pt}
\end{tabular}
\caption{Experimental results on Weibo and DRWeibo dataset.}
\label{tab:weibo}
\end{table*}

\begin{table*}[!h]
\centering
\begin{tabular}{ccccccccccc}
 \Xhline{1.0pt}
 \rowcolor{gray!20}
 ~ & \multicolumn{5}{c}{\textbf{Twitter15}} & \multicolumn{5}{c}{\textbf{Twitter16}}\\
 \cline{2-11}
 \rowcolor{gray!20}
 ~ & ~ & \textbf{N} & \textbf{F} & \textbf{T} & \textbf{U} & ~ & \textbf{N} & \textbf{F} & \textbf{T} & \textbf{U}\\
 \cline{3-6}
 \cline{8-11}
 \rowcolor{gray!20}
 \multirow{-3}{*}{\textbf{Method}} & \multirow{-2}{*}{\textbf{Acc.}} & \textbf{F1} & \textbf{F1} & \textbf{F1} & \textbf{F1} & \multirow{-2}{*}{\textbf{Acc.}} & \textbf{F1} & \textbf{F1} & \textbf{F1} & \textbf{F1}\\
 \hline
 PLAN & 0.819\small ±0.004 & 0.839 & 0.854 & 0.817 & 0.759 & 0.843\small ±0.005 & \textbf{0.855} & 0.851 & 0.858 & 0.805 \\
 BiGCN & 0.844\small ±0.005 & 0.856 & 0.844 & 0.863 & 0.809 & 0.880\small ±0.009 & 0.793 & 0.912 & 0.947 & 0.849 \\
 UDGCN & 0.840\small ±0.005 & 0.848 & 0.847 & 0.864 & 0.799 & 0.875\small ±0.009 & 0.783 & 0.902 & 0.954 & 0.839 \\
 GACL & 0.846\small ±0.007 & 0.859 & 0.845 & 0.866 & 0.812 & 0.891\small ±0.004 & 0.802 & 0.929 & 0.945 & 0.872 \\
 DDGCN & 0.835\small ±0.006 & 0.840 & 0.850 & 0.856 & 0.791 & 0.893\small ±0.004 & 0.807 & \textbf{0.931} & 0.946 & 0.871 \\
 \textbf{RAGCL} & \textbf{0.867\small ±0.005} & \textbf{0.891} & \textbf{0.867} & \textbf{0.869} & \textbf{0.835} & \textbf{0.905\small ±0.003} & 0.836 & 0.923 & \textbf{0.963} & \textbf{0.882} \\
 \Xhline{1.0pt}
\end{tabular}
\caption{Experimental results on Twitter15 and Twitter16 dataset.}
\label{tab:twitter}
\end{table*}

\subsection{Ablation Study}

We conducted a series of ablation experiments to verify the influence of different factors on the model performance.

\subsubsection{Unresponded 1-level Replies}

\begin{figure}[!h]
  \centering
  \includegraphics[width=8cm]{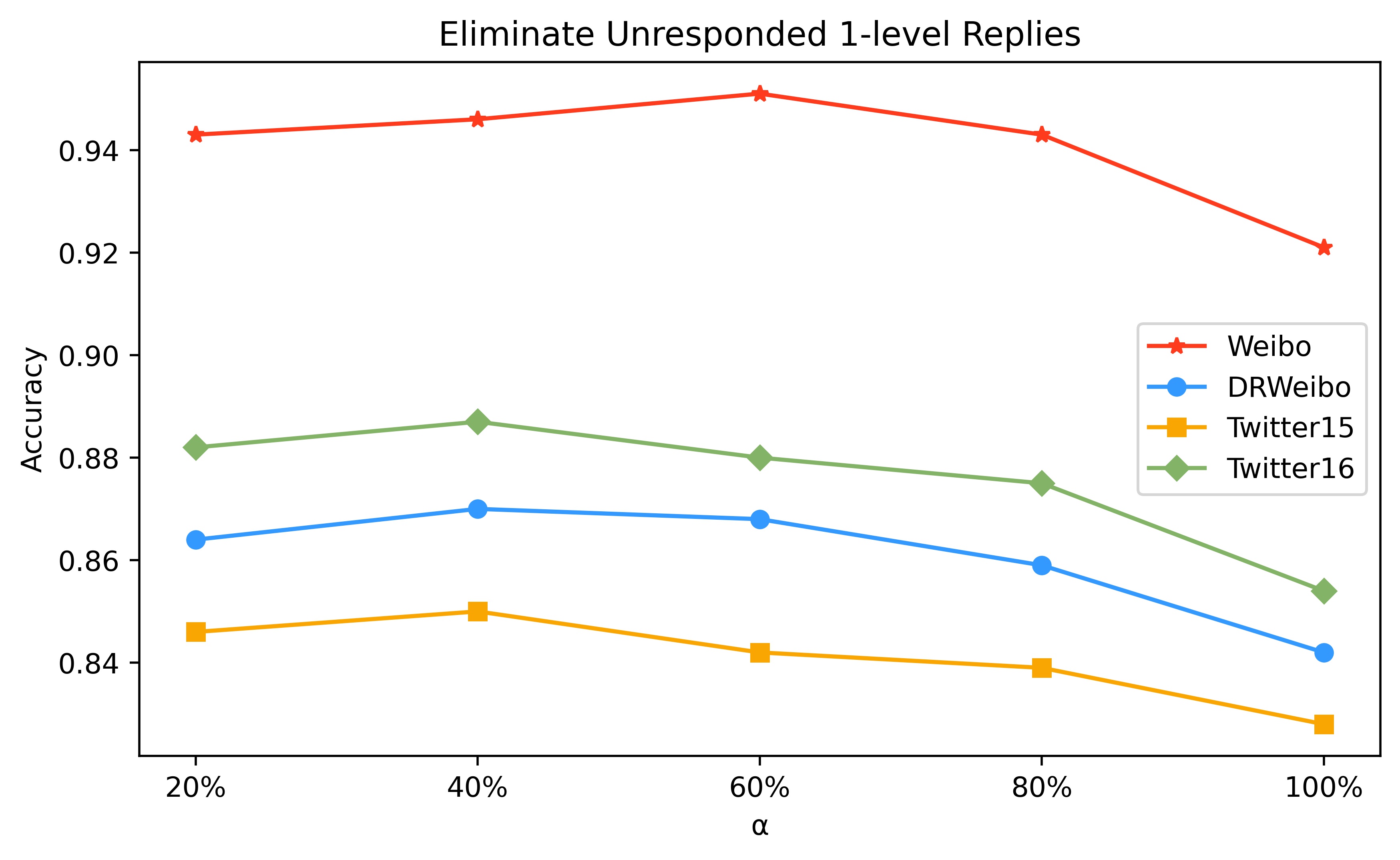}
  \caption{The influence of unresponded 1-level replies.}
  \label{fig:ab}
\end{figure}

In order to validate the impact of unresponded 1-level replies within RPTs, we conducted experiments on the four datasets depicted in Figure~\ref{fig:ab}. We eliminated $\alpha \%$ of unresponded 1-level replies in each RPT, subsequently utilizing BiGCN \cite{bigcn} for classification. With the increase of $\alpha$, it is observed that the model performance remains steady, or even improves to some extent. This indicates that these unresponded 1-level replies, as we previously conjectured, have less significance or may even serve as noise within rumor classification process, thus RAGCL is justified in dropping them.

\subsubsection{Data Augmentation Combinations}

Table~\ref{tab:aug} presents the impact of different data augmentation combinations, where we report the accuracy for each dataset. The experimental results show that using attribute masking in Chinese datasets (Weibo and DRWeibo) will reduce the model performance. For English datasets, various data augmentation combinations have minimal effect on the results. Different data augmentation combinations all achieve significant performance gains over using only GCN for supervised classification without applying contrastive loss. Furthermore, the results also indicate that adaptive data augmentation outperforms random data augmentation, providing further validation of the reliability of our theory.

\begin{table*}[!h]
\centering
\begin{tabular}{cccccc}
 \Xhline{1.0pt}
 \rowcolor{gray!20}
 \textbf{Aug1} & \textbf{Aug2} & \textbf{Weibo} & \textbf{DRWeibo} & \textbf{Twitter15} & \textbf{Twitter16}\\
 \hline
 - & - & 0.927 & 0.844 & 0.822 & 0.846 \\
 Node Dropping (random) & Attr Masking (random) & 0.940 & 0.861 & 0.837 & 0.865\\
 \hdashline
 Node Dropping & Attr Masking & 0.953 & 0.892 & \textbf{0.867} & 0.896\\
 Node Dropping & Edge Dropping & \textbf{0.962} & \textbf{0.894} & 0.864 & 0.902 \\
 Attr Masking & Edge Dropping & 0.952 & 0.888 & 0.864 & \textbf{0.905}\\
 \Xhline{1.0pt}
\end{tabular}
\caption{The influence of combinations of data augmentation.}
\label{tab:aug}
\end{table*}

\subsubsection{Node Centrality Measures}

We conducted the experiments in Table~\ref{tab:nc} to explore the influence of different node centrality measures. We report the accuracy that RAGCL achieves with different node centrality measures and the average time cost (in seconds) to calculate each RPT centrality. 
Degree centrality can be calculated rapidly, thus yielding efficient determination of node centrality within sizable datasets. However, its exclusive focus on edge number fails to satisfy Principle 3, thereby highlighting a limitation of degree centrality. For instance, a parent node and one of its children possessing identical reply counts will be assigned the same centrality. In fact, degree centrality also achieves relatively poor performance. Betweenness centrality aligns well with the three principles. For RPTs, the betweenness centrality is a very intuitive index to measure the importance of nodes. A node with numerous successor nodes will have many shortest paths traversing through it, leading to a correspondingly elevated betweenness centrality. However, the computation of betweenness centrality is more complex and time-intensive than the other measures. PageRank centrality, on the other hand, not only aligns well with the basic principles but also benefits from a relatively swift calculation process, making it more conducive to RAGCL's training phase.
We also examined the effect of eigenvector centrality, Katz centrality, and closeness centrality to verify the validity of our three guiding principles. Given their individual characteristics, these measures fail to meet Principle 2 and 3, resulting in subpar performance. Additionally, their computational complexity is relatively high. Therefore, we do not recommend using these centrality measures in RAGCL.

\begin{table}[!h]
\centering
\begin{tabular}{cccccc}
 \Xhline{1.0pt}
 \rowcolor{gray!20}
 ~ & ~ & \multicolumn{2}{c}{\textbf{Weibo}} & \multicolumn{2}{c}{\textbf{Twitter15}}\\
 \cline{3-6}
 \rowcolor{gray!20}
 \multirow{-2}{*}{\textbf{Centrality}} & \multirow{-2}{*}{\textbf{$\mathbf{T(n)}$}} & \textbf{Acc.} & \textbf{Time} & \textbf{Acc.} & \textbf{Time} \\
 \hline
 Degree & $O(1)$ & 0.953 & \textbf{0.82} & 0.860 & \textbf{0.13} \\
 Betweenness & $O(n^3)$ & 0.958 & 8.12 &  \textbf{0.867} & 1.34 \\
 PageRank & $O(n)$ & \textbf{0.962} & 1.37 &  0.865 & 0.22 \\
 \hdashline
 Eigenvector & $O(n^3)$ & 0.939 & 9.23 & 0.850 & 1.62 \\
 Katz & $O(n^3)$ & 0.943 & 9.37 & 0.849 & 1.67 \\
 Closeness & $O(n^3)$ & 0.935 & 7.72 &  0.841 & 1.44 \\
 \Xhline{1.0pt}
\end{tabular}
\caption{The influence of node centrality measures.}
\label{tab:nc}
\end{table}

\subsubsection{Graph Direction}

\begin{figure}[h]
  \centering
  \subfigure[Weibo]{\includegraphics[width=0.23\textwidth]{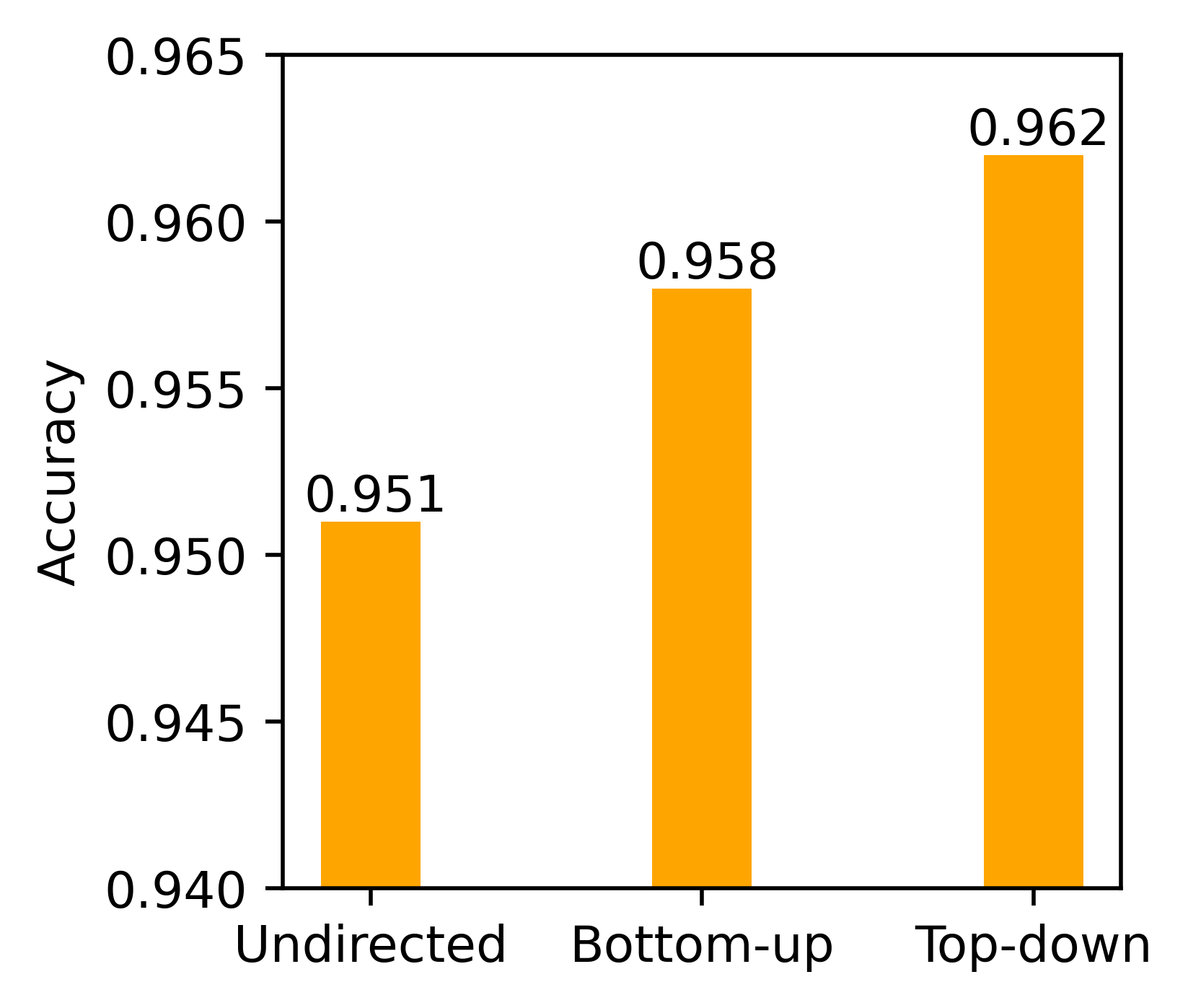}}
  \subfigure[Twitter15]{\includegraphics[width=0.23\textwidth]{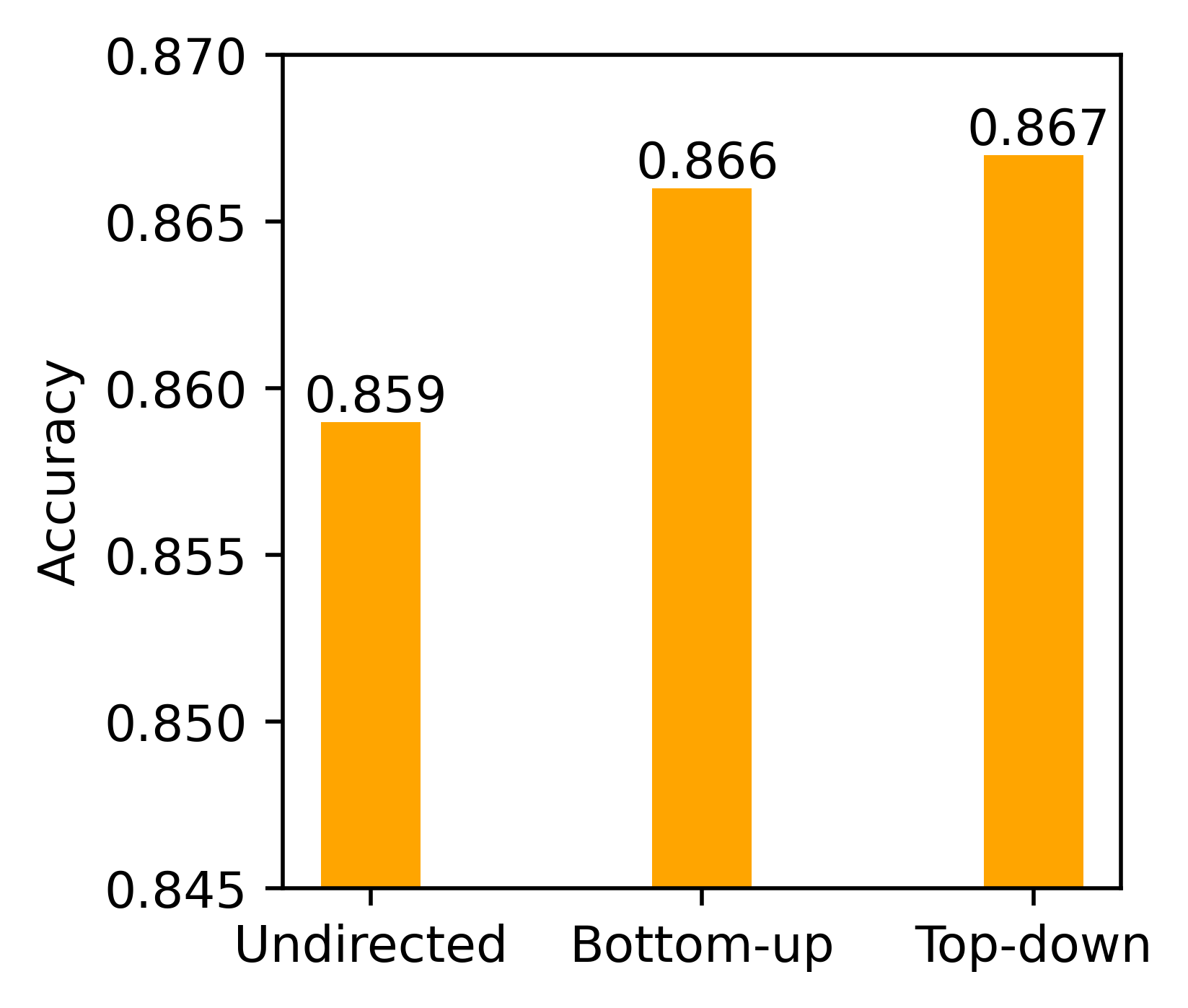}}
  \caption{The impact of information flow direction.}
  \label{fig:direction}
\end{figure}

RAGCL is compatible with top-down and bottom-up directed graphs as well as undirected graphs. We investigated the impact of different types of graph in Figure~\ref{fig:direction}. The results show that using undirected graphs leads to a performance decline. This could be due to the fact that during the forward propagation process of GNNs, densely connected nodes at the root node will see each other in their neighboring field of view. These nodes mutually aggregate each other's information, ultimately resulting in a loss of node feature uniqueness, causing an over-smoothing problem \cite{oversm1,oversm2,oversm3}. On the other hand, top-down and bottom-up directed graphs are able to effectively block excessive information flow between nodes at the root node.

\section{Conclusion}

This study introduces RAGCL, an adaptive graph contrastive learning method specifically for rumor detection. By taking into consideration the structural characteristics of RPTs, we propose three adaptive data augmentation methods based on node centrality and provide guiding principles for designing these methods. Our experimental results demonstrate that RAGCL surpasses current SOTA methods on all datasets, showcasing its superior performance.

\section{Acknowledgments}

This work is supported in part by the National Key R\&D Program of China (2018AAA0100302) and the National Natural Science Foundation of China (61876016).

\bibliography{aaai24}

\end{document}